\documentstyle[aps,prb]{revtex}

\newcommand{\beq}{\begin{equation}}
\newcommand{\eeq}{\end{equation}}

\begin{document}

\author{J.O. Fj{\ae}restad,$^{1}$ A. Sudb{\o}$^{1}$ and A. Luther$^{2}$}

\address{$^{1}$ Department of Physics, Norwegian University of Science and 
Technology, N-7034 Trondheim, Norway. \\
$^{2}$ Nordisk Institut for Teoretisk Fysik (NORDITA), Blegdamsvej 17, 
DK-2100 Copenhagen {\O}, Denmark.}

\date{\today}

\title{\bf Correlation functions for a two-dimensional electron system
\\with bosonic interactions and a square Fermi surface}

\maketitle

\begin{abstract}
We calculate zero-temperature correlation functions for a model of 2D
interacting electrons with short-range interactions and a square Fermi
surface. The model was arrived at by mapping electronic states near a
square Fermi surface with Hubbard-like interactions onto
one-dimensional quantum chains, retaining terms which can be written
in terms of bosonic density operators.  Interactions between
orthogonal chains, corresponding to orthogonal faces of the square
Fermi surface, are neglected. The correlation functions become sums of
Luttinger-type correlation functions due to the bosonic
model. However, the correlation function exponents differ in form from
those of the Luttinger model. As a consequence, the simple scaling
relations found to exist between the Luttinger model exponents, do not
carry over to the leading exponents of our model.  We find that for
repulsive effective interactions, charge-density wave/spin-density
wave instabilities are dominant.  We do not consider $d$-wave
instabilities here.
\end{abstract}



\section{INTRODUCTION}
\label{intro}

In recent years, research in theoretical condensed matter physics has
focused intensely on the possible breakdown of Landau's Fermi-liquid
theory in strongly correlated electron systems with spatial dimension
larger than one.  This has been motivated by Anderson's observation
\cite{anderson} that the normal state of the high-$T_c$
superconductors has characteristics which seem irreconcilable with the
predictions of Fermi liquid theory.\cite{varma} It is well known that
one-dimensional (1D) systems of strongly correlated electrons with
gapless excitations belong to the socalled Luttinger liquid
universality class, \cite{haldane,voit} but an analogous rigorous
theoretical foundation for non-Fermi liquid physics in two-dimensional
(2D) systems is still lacking.

A number of authors have treated the 2D interacting electron gas with
a circular Fermi surface using bosonization and/or
renormalization-group techniques.\cite{houghton93,fradkin93,shankar94}
Resummations of perturbation theory have also been carried
out,\cite{castellani94} including the case of a square Fermi surface
within the parquet-approximation.\cite{dzyal95} It is invariably found
that the Luttinger-liquid fixed point is unstable in more than one
spatial dimension. The underlying physics is that, due to the
restricted kinematics on the Fermi surface in one spatial dimension,
forward scattering is singular. In higher dimensions, small-angle
scattering appears to avoid such a singularity, unless the
interactions are very long-ranged.  \cite{stamp92} On the other hand,
for a nested or nearly-nested Fermi surface such as is observed in the
high-$T_c$ cuprates,\cite{shen93} the phase-space for small-angle
scattering is suppressed. Under such circumstances, it appears that a
breakdown of Fermi-liquid theory may be feasible even with regular
interactions.\cite{luther94} It should be noted that approaches based
on, for instance, resummation of perturbation series, such as the
parquet-approximation of Ref. \onlinecite{dzyal95}, are expected to be
reliable for not too strong correlations.  Alternatively, one could
consider a regime of interaction-parameters where correlation effects
give rise to qualitatively new phenomena that should be incorporated
even in a zeroth order approximation to the problem. This would then
complement, for instance, the work of Ref. \onlinecite{dzyal95}.

Recently, a 2D system of electrons on a square lattice with nearest
neighbour hopping and Hubbard-like interactions was considered, when
the Fermi surface is a square even away from
half-filling.\cite{luther94} The necessary requirement for this to be
a valid starting point is that the interactions of the underlying
lattice model must be strong enough. This approach therefore does not
have a free-fermion limit; the zeroth order Hamiltonian to the problem
is rather that of free bosons. In Ref. \onlinecite{luther94}, the problem
was mapped onto two sets of one-dimensional chains, one set for each
axis of the square Fermi surface. The kinetic energy and the
interaction were separated into terms which could be written in terms
of bosonic density operators for the chains, and terms of non-bosonic
form. The bosonic Hamiltonian was then diagonalized, assuming no
interactions between orthogonal chains, and the single particle
Green's functions were evaluated, with a resulting Luttinger-liquid
form.  The non-bosonic terms in the Hubbard-like interactions were
classified as relevant or irrelevant with respect to the solution of
the bosonic Hamiltonian. Discarding irrelevant operators, the
Hamiltonian for a particular case of sufficiently strong repulsive
interactions was solved, treating the relevant operators as
perturbations on the bosonic solution.

The $N$-component one-dimensional electron gas has very recently been
treated using bosonization,\cite{emery98} obtaining the operator
dimensions for a number of non-bosonic terms that could appear in the
Hamiltonian.\cite{luther94a} One particular realization of such a
system is the $N$-chain Hubbard model. Given the fact that we neglect
interactions between orthogonal chains in our model, the system we
effectively study has much in common with that of
Ref. \onlinecite{emery98}. However, our work differs in one important
respect, namely that we consider correlation functions of fermion
operators that are non-local in chain indices.

The minimal microscopic model appropriate for the description of
high-$T_c$ superconductivity is not yet agreed upon. Much work has
been done for Hubbard-like models, also for the square Fermi surface
case.  \cite{mattis87,hlubina94,kwon97,abreu97,zanchi96} Recently,
claims were however made that non-existence of $d_{x^2-y^2}$-wave
superconductivity in the 2D Hubbard model had been rigorously proved
at any finite temperature, and under certain circumstances even at
zero temperature.\cite{su98} If that were to be true, the single-band
Hubbard model is not likely to be a fruitful microscopic starting
point for explaining the phenomenology of the high-$T_c$ cuprates,
where there is now broad consensus on a dominant
$d_{x^2-y^2}$-component of the order parameter. In this situation, the
examination of the properties of various hypothetical effective
theories may give useful information about what directions one should
take in solving this exceedingly difficult problem.

In this paper, we investigate the ground state properties of the
bosonic Hamiltonian in more detail, through the calculation and
analysis of explicit expressions for various ground state correlation
functions, which enables us to identify the dominant divergent
susceptibilities. We calculate the correlation functions for $2k_F$
charge-density wave (CDW), $2k_F$ spin-density wave (SDW), $s$-wave
singlet (SS), and triplet (TS) superconductivity fluctuations. The
similarities to the results obtained for the Luttinger model are: (i)
the separate terms making up the correlation functions have
Luttinger-model form, due to the bosonic form of our model, (ii) there
is a degeneracy between CDW and SDW fluctuations, and between SS and
TS fluctuations, (iii) for effective repulsive interactions, the
CDW/SDW instabilities are dominant.  However, because the expressions
for our correlation function {\em exponents} differ in form from those
of the Luttinger model, the simple scaling relations valid for the
exponents of the Luttinger model do not hold for our model.

This paper is organized as follows. In Sec. \ref{model}, we describe
the fundamentals of the model under consideration. In Sec. \ref{corr},
various ground state correlation functions for the model are
calculated, and their exponents are discussed.  Sec. \ref{disc}
summarizes the obtained results.

\section{THE MODEL}
\label{model}

In this section, we give an introduction to the model under
consideration, and establish the notation that will be used in the
rest of the paper.  For a more detailed discussion we refer the reader
to Ref.  \onlinecite{luther94}.

\subsection{The kinetic energy. Mapping to chains}

We consider a 2D electron system on a square lattice with 
lattice constant $a$ and $N$ lattice points in each direction (although we
will not indicate it explicitly, the thermodynamic limit 
$N\to\infty$ will always be implied at the end of our calculations). The 
kinetic energy operator is
\beq
H_{\mbox{\scriptsize kin\normalsize}}=-t\sum_{\langle  ij \rangle s}
c^{\dagger}_{is}c_{js},
\label{kin}
\eeq
where $c^{\dagger}_{is}$ and $c_{is}$ are creation and annihilation operators 
for an electron with spin $s$ at site $i$. The sums over $i$ and $j$ are 
restricted to nearest neighbours, where $t$ is the hopping matrix element. 
We introduce coordinates $(x,y)$ in real space and $(k_x,k_y)$ in reciprocal 
space, where the coordinate axes are rotated +45 degrees with respect to the 
primitive vectors connecting nearest-neighbour sites. The kinetic energy 
operator can then be written
\beq
H_{\mbox{\scriptsize kin\normalsize}}=-4t\sum_{\vec{k}s}\cos(k_x a/\sqrt{2})
\cos(k_y a/\sqrt{2})c^{\dagger}_{\vec{k}s}c_{\vec{k}s},
\label{kink}
\eeq
where the $\vec{k}$ summation is over the 1st Brillouin zone. For a 
half-filled band, the Fermi surface is a square, with faces defined by 
$k_x$ or $k_y$ equal to $\pm k_F$ where $k_F\equiv\pi/\sqrt{2}a$, giving a
vanishing Fermi energy. 

Since we are interested in the low-energy physics, the kinetic 
energy is linearized in the direction perpendicular to each of the four faces 
of the square Fermi surface. Consider the two parallel faces defined by 
$k_x=\alpha k_F$, $-k_F\leq k_y \leq k_F$, where $\alpha=\pm 1$. These faces 
will from now on be referred to as the '+' and '-' faces, 
respectively. The linearized kinetic energy dispersion near face $\alpha$ is
\beq
E_{\mbox{\scriptsize kin\normalsize},\alpha}(\vec{k})=v_F(k_y)(\alpha k_x-k_F),
\eeq
where $v_F(k_y)=2\sqrt{2}\,ta\,\cos(k_y a/\sqrt{2})$. This form is 
reminiscent of the dispersion for the Luttinger model, 
except that in our case, the dispersion is only valid within a finite 
region ($-k_F\leq k_y\leq k_F$), and the Fermi velocity $v_F(k_y)$ depends on 
the momentum parallel to the face, vanishing on the edges of the face. We 
also define the field operator for excitations near face $\alpha$ as
\beq
\psi_{\alpha s}(x,y)=\frac{1}{N}\sum_{k_x,k_y}e^{ik_x x+ik_y y}
c_{\alpha s}(k_x),
\label{orop}
\eeq
where the summation over $k_y$ is between $-k_F$ and $k_F$, and the summation
over $k_x$ should be cut off for $\alpha k_x\ll k_F$. 
Similar considerations and definitions can be made for the two other 
faces of the square Fermi surface.

We now consider a two-point function like $\langle \psi^{\dagger}(x,y,t)
\psi(0,0,0)\rangle$, where the average is taken with respect to the ground
state of the kinetic energy operator. The total field operator is written
as a sum of the four face field operators of the type given in Eq. 
(\ref{orop}), making the two-point function above a sum of four terms, one
from each face.
The contribution from face $\alpha$ will be $\langle \psi^{\dagger}_
{\alpha}(x,y,t)\psi_{\alpha}(0,0,0)\rangle$. It can be shown that 
the result for this correlator is exactly reproduced by the following
mapping of the field operator,
\beq
\psi_{\alpha s}(x,y)\rightarrow\sqrt{\frac{\pi}{k_F}}\sum_{l'=0}^{2N-1}g(l-l')
\psi_{\alpha l' s}(x)
\label{mapping}
\eeq
with the resulting mapping of the unperturbed Hamiltonian,
\beq
H_{\mbox{\scriptsize kin\normalsize},\alpha}\rightarrow 
\sum_s \sum_{l,l'}t_{l,l'}\int_0^L dx 
\psi^{\dagger}_{\alpha l s}(x)\left(
\frac{\alpha}{i}\frac{\partial}{\partial x}-k_F\right)\psi_{\alpha l' s}(x),
\label{kinchains}
\eeq
where
\beq
t_{l,l'}=\frac{1}{2N}\sum_{n_y=-N/2}^{N/2-1}v_{F}(k_y)e^{ip_y(l-l')}.
\label{tll}
\eeq
Here the integer $l=\sqrt{2}y/a\in[0,2N-1]$, $k_y=2\pi n_y/\sqrt{2}Na$, 
$p_y\equiv\pi n_y/N$, so that $k_y y=p_y l$. Furthermore,
\beq
g(l)=\frac{1}{2N}\sum_{n_y=-N/2}^{N/2-1}e^{ip_y l}
=\frac{(-i)^l}{2N}\frac{1-(-1)^l}{1-e^{i\pi l/N}}
\eeq
satisfying $\sum_{l_1}g(l-l_1)g(l_1-l')=g(l-l')$. 
Letting $N\to\infty$ for fixed $l$, we
obtain
\beq
g(l)=\frac{a}{\sqrt{2}}\int_{-k_F}^{k_F}\frac{dk_y}{2\pi}e^{-ik_y (y)}
=\frac{1}{2}\delta_{l,0}+\frac{\sin[\pi l/2]}
{\pi l}(1-\delta_{l,0}).
\label{ginf}
\eeq

The operators $\psi_{\alpha l s}(x)$ describe right-moving ($\alpha=1$) and
left-moving ($\alpha=-1$) electrons living on fictitious one-dimensional 
chains (labeled by $l$) of length $L=Na$. In terms of operators creating 
electrons with definite momenta, these chain field operators are given by
\beq
\psi_{\alpha l s}(x)=\frac{1}{\sqrt{L}}
\sum_{k_x}e^{ik_x x} ~ c_{\alpha l s}(k_x),
\eeq
where the normalization indicates that $x$ is now taken to be 
a continuous variable. The discrete wavevectors are given by $k_x=2\pi n_x/L$,
with integer $n_x$. Note that since $g(l-l')$ is generally non-zero for 
$l\neq l'$, the mapping (\ref{mapping}) is non-local.

We take the non-interacting system at half-filling with linearized
dispersion as our unperturbed problem. Perturbations, in the form of
interactions or operators taking the system away from half-filling,
can in principle be treated within the framework of many-body
perturbation theory, where their effect is expressed in terms of
integrals over time-ordered multiple-point correlation
functions. Using Wick's theorem, such correlation functions can be
expressed as sums of products of two-point functions. Since the latter
are correctly reproduced by the mapping (\ref{mapping}) (with
(\ref{kinchains}) governing the dynamics), the entire perturbation
series will be correctly generated.  This, together with the fact that
standard 1D bosonization can be applied to the chains, is what makes
the mapping useful.

Electron creation and annihilation operators belonging to {\em
different} faces and/or chains anticommute, by definition. Thus the
associated density operators $\rho_{\alpha l s}(n_x)=\sum_{k'_x}
:c^{\dagger}_{\alpha l s}(k'_x+k_x)c_{\alpha l s}(k'_x):$ ($:\cdots:$
means normal-ordering) commute. For the '$+/-$' faces this is
expressed by the commutator 
\beq [\rho_{\alpha l s}(n_x),\rho_{\alpha'
l' s'}(-n'_x)]= -\alpha n_x
\delta_{\alpha,\alpha'}\delta_{l,l'}\delta_{s,s'}\delta_{n_x,n'_x}.
\eeq 
This is simply the result for the usual single-chain case
generalized to the case of many chains, the only effect of the
generalization being the introduction of the factor
$\delta_{l,l'}$. The Fourier transformed operators \beq \rho_{\alpha
s}(\vec{n})=\sum_{l=0}^{2N-1}\rho_{\alpha l s}(n_x)e^{-ip_y l}, \eeq
where $\vec{n}=(n_x,n_y)$, then obey the commutation relations \beq
[\rho_{\alpha s}(\vec{n}),\rho_{\alpha' s'}(-\vec{n}')]= -2N\alpha n_x
\delta_{\alpha\alpha'}\delta_{ss'}\delta_{\vec{n},\vec{n}'}.
\label{commdensity}
\eeq

It was shown in Ref. \onlinecite{luther94} that in the presence of 
interactions,
the terms with $l'\neq l$ in (\ref{kinchains}) are irrelevant in the RG sense.
Therefore we only keep the diagonal part, which we define as 
$H^{d}_{\mbox{\scriptsize kin\normalsize},\alpha}$. This will be a sum of 
independent chains. Bosonization is performed for each chain, giving
\beq
H_{\mbox{\scriptsize kin\normalsize},\alpha}^{d}=
\frac{\pi v_0}{2NL}\sum_{\vec{n},s}
:\rho_{\alpha s}(\vec{n})\rho_{\alpha s}(-\vec{n}):,
\label{kinboson}
\eeq
where 
$v_0\equiv t_{ll}=2\sqrt{2}ta/\pi$, and the sum over $n_y$ is between $-N$ 
and $N-1$ (this will be the summation range for $n_y$ in the rest of the 
article as well).

Completely analogous considerations can be made for the two faces with
$k_y=\pm k_F$, the only difference being that the associated chains will then
lie along the $y$ direction, taken to be continuous, while the $x$ direction
is kept discrete. Boson density operators for perpendicular faces commute.

\subsection{Interactions. Solution of the bosonic Hamiltonian}

The form of the interactions which are included in the bosonic 
Hamiltonian can be extracted from the Hubbard model interaction, given by
$U\sum_{i}n_{i\uparrow}n_{i\downarrow}$, where $U$ is the on-site repulsion
and $n_{is}$ is the number operator at site $i$ with spin $s$. Writing the 
field operator as a sum of the field operators for the four faces of the 
square Fermi surface, the Hubbard interaction will contain terms which 
couple perpendicular faces and terms which only couple parallel faces.
The former terms are not included in the bosonic Hamiltonian, while from
the latter terms, which for the $'+/-'$ faces can be written 
\begin{eqnarray} & &
U\left(\frac{\pi}{k_F}\right)^2\sum_{x,y}\sum_{\stackrel{\alpha_1\alpha_2}
{\alpha_3\alpha_4}}
\sum_{\stackrel{l_1 l_2}{l_3 l_4}}g(l-l_1)g(l-l_2)g(l-l_3)g(l-l_4) 
 :\psi^{\dagger}_{\alpha_1 l_1 \uparrow}(x)
\psi_{\alpha_2 l_2 \uparrow}(x)
\psi^{\dagger}_{\alpha_3 l_3 \downarrow}(x)\psi_{\alpha_4 l_4 \downarrow}(x):,
\end{eqnarray}
we retain the bosonic contributions, 
which have the form of density-density interactions, obtained when 
$(\alpha_1, l_1)=(\alpha_2, l_2)$ and $(\alpha_3, l_3)=(\alpha_4, l_4)$. 
(Note that since perpendicular faces are decoupled in the bosonic Hamiltonian,
it is from now on sufficient to focus on one of the two sets of parallel
faces, and we choose the $'+/-'$ faces.) 
The bosonic contributions can be written
\beq
\frac{Ua}{NL}\sum_{\alpha\alpha'}\sum_{\vec{n}}f^2(n_y)
:\rho_{\alpha\uparrow}(\vec{n})\rho_{\alpha'\downarrow}(-\vec{n}):,
\label{orboseint}
\eeq
where $f(n_y)=(1-|n_y|/N)/2$ is the Fourier transform of $g^2(l)$.
Introducing charge ($\nu=\rho$) and spin ($\nu=\sigma$) operators as
($\eta_{\rho}=1$, $\eta_{\sigma}=-1$)
\beq
\sqrt{2}\nu_{\alpha}(\vec{n}) = \rho_{\alpha\uparrow}(\vec{n})
+\eta_{\nu} \rho_{\alpha\downarrow}(\vec{n}),
\label{chspin}
\eeq
with commutation relations $(\nu,\nu'=\rho,\sigma)$
\beq
[\nu_{\alpha}(\vec{n}),\nu'_{\alpha'}(-\vec{n}')]=-2N\alpha n_x
\delta_{\nu\nu'}\delta_{\alpha\alpha'}\delta_{\vec{n},\vec{n}'},
\eeq
the most general interaction of form (\ref{orboseint}) can be written
on bosonized form as follows:
\beq
H_{B} \equiv  \frac{a}{2NL}\sum_{\nu=\rho,\sigma}
\sum_{\alpha\alpha'}\sum_{\vec{n}}
f^2(n_y)\eta_{\nu}U^{(\nu)}_{\alpha\alpha'}(n_x)
:\nu_{\alpha}(\vec{n})\nu_{\alpha'}(-\vec{n}): 
\eeq
where we have allowed for different coupling functions
$U_{\alpha\alpha'}^{(\nu)}(n_x)$ for the different scattering processes. Hence,
this bosonic interaction is a bit more general than that extracted directly
from the Hubbard model. In analogy with the 1D case, a cutoff on the momentum 
transfer in the $x$ direction will be needed in order to get a well-defined 
solution; this makes the couplings $n_x$-dependent. Symmetry
considerations dictate that there can be at most four different coupling
functions: 
$U^{(\rho)}_{++}$, $U^{(\rho)}_{+-}$, $U^{(\sigma)}_{++}$, and $U^{(\sigma)}
_{+-}$. 

The kinetic energy $H_0\equiv \sum_{\alpha}H^{d}_{\mbox{\scriptsize kin
\normalsize},\alpha}$ can also be 
written in terms of charge and spin operators,
\beq
H_0=\frac{\pi v_0}{2NL}\sum_{\nu=\rho,\sigma}\sum_{\alpha}\sum_{\vec{n}}
:\nu_{\alpha}(\vec{n})\nu_{\alpha}(-\vec{n}):.
\eeq
The Hamiltonian $H_0+H_B$, which shows spin-charge separation, will be 
referred to as the bosonic Hamiltonian. It may be diagonalized 
separately in the spin and charge sectors, using the canonical transformation 
$\exp{(\sum_{\nu}S_{\nu})}(H_0+H_B)\exp{(-\sum_{\nu}S_{\nu})}\equiv H_D$, where
\beq
S_{\nu}=\frac{1}{2N}\sum_{\vec{n}}\frac{\xi_{\nu}(\vec{n})}{n_x}
\nu_{+}(\vec{n})\nu_{-}(-\vec{n}),
\label{cantr}
\eeq
and $\xi_{\nu}(\vec{n})$ is chosen so that the
off-diagonal terms in the transformed Hamiltonian $H_D$ vanish. This gives
\beq
H_D=\frac{\pi}{2NL}\sum_{\nu}\sum_{\alpha}\sum_{\vec{n}}
v_{\nu}(\vec{n}):\nu_{\alpha}(\vec{n})\nu_{\alpha}(-\vec{n}):,
\label{hd}
\eeq
where the parameters $\xi_{\nu}(\vec{n})$ and the renormalized velocities
$v_{\nu}(\vec{n})$ are given by 
\begin{eqnarray}
\exp{[2\xi_{\nu}(\vec{n})]}&=&\sqrt{\frac{1+a\eta_{\nu}[U^{(\nu)}_{++}(n_x)+
U^{(\nu)}_{+-}(n_x)]f^2(n_y)/\pi v_0}{1+a\eta_{\nu}[U^{(\nu)}_{++}(n_x)
-U^{(\nu)}_{+-}(n_x)]f^2(n_y)/\pi v_0}}, \label{xiformula}\\
\frac{v_{\nu}(\vec{n})}{v_0}&=&\sqrt{\left(1+a\eta_{\nu}U^{(\nu)}_{++}
(n_x)f^2(n_y)/\pi v_0\right)^2-\left(a\eta_{\nu}U^{(\nu)}_{+-}(n_x)
f^2(n_y)/\pi v_0
\right)^2}.
\label{vformula}
\end{eqnarray}
Terms in the microscopic Hamiltonian which are not of the type
included in $H_D$ will, if they are irrelevant, only lead to
modifications in the numerical values of the coupling functions,
leaving the structure of the low-energy theory unaffected. In this
case, it is these 'effective' function values that should enter in the
formulae above.  As already pointed out below Eq. (\ref{commdensity}),
the non-diagonal terms (i.e. $l\neq l'$) in the kinetic energy
(\ref{kinchains}) are irrelevant perturbations with respect to
$H_D$.\cite{luther94} We emphasize that for the case of the 2D
repulsive Hubbard model, Umklapp scattering, which is of non-bosonic
form, is a relevant interaction.\cite{luther94} The general philosophy
of the approach presented here, is to view $H_D$ as a zero-order
Hamiltonian for a perturbation treatment of the relevant non-bosonic
interactions generated by the underlying microscopic lattice
Hamiltonian (which may be different from the Hubbard model).

From the high-$T_c$ point of view, the interesting case is doping away
from half-filling. In Ref. \onlinecite{luther94}, it was shown that for
sufficiently strong interactions, the only effect of the operator causing
deviations from half-filling was to shift the value of the Fermi wave vector,
not changing the square shape of the Fermi surface. In the remainder of the
paper, we will assume that we are in this strong-interaction regime with a
square Fermi surface.

\section{CORRELATION FUNCTIONS}
\label{corr}

We now turn to the calculation of correlation functions for various 
fluctuations of interest in the ground state $|\rangle$ of the bosonic 
Hamiltonian $H_0+H_B$. The dominant instability is identified by
the correlation function with the slowest space-time decay, thus having
a generalized susceptibility with the strongest divergence as $T\to 0$.

\subsection{Definitions of correlation functions. 
Calculation of their $y$-dependence}

The generic form of the correlation functions we consider is 
\beq
R^{>}(\vec{r},t)=-i\langle | O(\vec{r},t)O^{\dagger}(0,0)|\rangle,
\label{corrfuncdef}
\eeq
where $O(\vec{r},t)$ is the operator for the fluctuation under consideration. 
We will consider the following fluctuations: $2k_F$ charge-density wave (CDW), 
$2k_F$ spin-density wave (SDW), $s$-wave singlet superconductivity (SS), and 
triplet superconductivity (TS). The SDW operator has 3 spatial components, 
and the TS operator has 3 components corresponing to the total spin in the $z$
direction. Assuming spin-rotation invariance, 
the 3 associated correlation functions should be equal, although the 
expressions will be formally different due to the abelian bosonization used 
here.\cite{voit} 
The definitions of the operators $O$ are simple generalizations of the 
corresponding 1D definitions:\cite{solyom,voit} 
\begin{eqnarray}
O_{CDW}(\vec{r}) &=& \sum_s \psi^{\dagger}_{+,s}(\vec{r})
\psi_{-,s}(\vec{r}),\\
O_{SDW,x}(\vec{r})&=&\sum_s \psi^{\dagger}_{+,s}(\vec{r})
\psi_{-,-s}(\vec{r}),\\
O_{SDW,z}(\vec{r})&=&\sum_s s\,\psi^{\dagger}_{+,s}(\vec{r})
\psi_{-,s}(\vec{r}),\\
O_{SS}(\vec{r}) &=& \frac{1}{\sqrt{2}}\sum_s 
\psi^{\dagger}_{+,s}(\vec{r})
\psi^{\dagger}_{-,-s}(\vec{r}), \\
O_{TS,+1}(\vec{r}) &=& \psi^{\dagger}_{+,\uparrow}(\vec{r})
\psi^{\dagger}_{-,\uparrow}(\vec{r}).
\end{eqnarray}
We have included the definitions of two of the components of the SDW operator,
since the equality of the associated correlation functions will give a 
condition for determining the parameter $\xi_{\sigma}(\vec{n})$.

Introducing the mapping (\ref{mapping}), and observing that the bosonic 
Hamiltonian conserves the number of electrons with a given spin on a given 
face and a given chain, it is clear that each non-zero term in the 
correlation functions has the form (discarding prefactors)
\begin{eqnarray}
F_{\beta,s_1 s_2}(l_1-l_2, x,t) \equiv
\langle | \psi^{\dagger}_{+,l_1,s_1}(x,t)\psi^{\beta}_{-,l_2,s_2}(x,t)
\psi^{-\beta}_{-,l_2,s_2}(0,0)\psi_{+,l_1,s_1}(0,0)|\rangle,
\end{eqnarray}
where $\beta=+1$ $(-1)$ indicates an annihilation (creation) operator. 
Here we have anticipated that $F$ is a function only of the product of the 
spins, and of the separation between the chains. In terms of $F$, the 
correlation functions are given by
\begin{eqnarray}
R^{>}_{CDW}(\vec{r},t) &\propto & \sum_{l_1=0}^{2N-1}\sum_{l_2=0}^{2N-1}
H_{l}(l_1,l_2)
F_{++}(l_1-l_2,x,t), \label{rcdw}\\
R^{>}_{SDW,x}(\vec{r},t) &\propto & \sum_{l_1=0}^{2N-1}\sum_{l_2=0}^{2N-1}
H_{l}(l_1,l_2)
F_{+-}(l_1-l_2,x,t), \label{rsdwx}\\
R^{>}_{SDW,z}(\vec{r},t) &\propto & \sum_{l_1=0}^{2N-1}\sum_{l_2=0}^{2N-1}
H_{l}(l_1,l_2)
F_{++}(l_1-l_2,x,t), \label{rsdwz}\\ 
R^{>}_{SS}(\vec{r},t) &\propto & \sum_{l_1=0}^{2N-1}\sum_{l_2=0}^{2N-1}
H_{l}(l_1,l_2)
F_{--}(l_1-l_2,x,t), \label{rss}\\
R^{>}_{TS,+1}(\vec{r},t) &\propto & \sum_{l_1=0}^{2N-1}\sum_{l_2=0}^{2N-1}
H_{l}(l_1,l_2)
F_{-+}(l_1-l_2,x,t),
\label{rts}
\end{eqnarray}
where $H_{l}(l_1,l_2)\equiv g(-l_1)g(l-l_1)g(-l_2)g(l-l_2)$. It is seen
already at this point that in this model, the CDW and SDW correlation 
functions are equal. 

In (\ref{rcdw})-(\ref{rts}) we first sum over all terms with 
a fixed value of $l_0\equiv l_1-l_2$, and then sum over the appropriate 
values of $l_0$. 
Using $H_{l}(l_1,l_2)=H_{l}(l_2,l_1)$ and $F(l_0,x,t)=F(-l_0,x,t)$ 
(the latter property is established in the
next subsection) we obtain
\beq
R(\vec{r},t)\propto K(l,0)F(0,x,t)+2\sum_{l_0=1}^{2N-1}K(l,l_0)F(l_0,x,t),
\label{RitoKF}
\eeq
where we have defined
\beq
K(l,l_0)=\sum_{l_1=l_0}^{2N-1}H_{l}(l_1,l_1-l_0).
\label{Kll0}
\eeq
For $l_0\geq 1$ it is convenient to write 
$K(l,l_0)=K_{+}(l,l_0)-K_{-}(l,l_0)$, where
\begin{eqnarray}
K_{+}(l,l_0) &=& \sum_{l_1=0}^{2N-1} H_{l}(l_1,l_1-l_0), \label{K+ll0def}\\
K_{-}(l,l_0) &=& \sum_{l_1=0}^{l_0-1}H_{l}(l_1,l_1-l_0). \label{K-ll0def}
\end{eqnarray}
In the calculation of $K_{+}$, the summation over $l_1$ must be done
before the $N\to\infty$ limit is taken.  This is because the
expressions (\ref{ginf}) are derived by assuming that $l$ is fixed,
and therefore finite, so that $l/N\to 0$ as $N\to\infty$. They are
therefore not correct when $l$ is of order $N$, and such values of
$l_1$ are indeed included in the sum in $K_{+}$.

The calculations of $K_{+}$ and $K_{-}$ are rather lengthy, 
so we only give the final results 
($\bar{\delta}_{l,l'}\equiv 1-\delta_{l,l'}$):
\begin{eqnarray}
K_{+}(l,l_0) &=& \delta_{l,0}\delta_{l_0,0}\frac{1}{12}
+\delta_{l,0}\bar{\delta}_{l_0,0}\frac{1}{2\pi^2 l_0^2}
+\delta_{l_0,0}\bar{\delta}_{\l,0}\frac{1}{2\pi^2 l^2}
+\bar{\delta}_{l,0}\delta_{l,l_0}
\frac{(-1)^{l+1}}{4\pi^2 l^2},\label{K+ll0res}\\
K_{-}(l,l_0) &=& \left(\frac{\sin(\pi l_0/2)}{2\pi l_0}\right)^2 \delta_{l,0}+
\frac{\cos^2(\pi l/2)\cos^2(\pi l_0/2)}{\pi^4}\nonumber \\ &\times &
\left\{\bar{\delta}_{l,0}\bar{\delta}_{l,l_0}\left[
\frac{\psi(\frac{1+l}{2})-\psi(\frac{1+l_0}{2})
+\psi(\frac{1}{2})}{l_0(l+l_0)(l-l_0)}-\frac{1}{2l l_0}
\left(\frac{\psi(\frac{1+l+l_0}{2})}{l+l_0}
+\frac{\psi(\frac{1+l-l_0}{2})}{l-l_0}\right)\right]\right. \nonumber \\ &+&
\left.\delta_{l,0}\left[\frac{2\left(\psi\left(\frac{1+l_0}{2}\right)
-\psi\left(\frac{1}{2}\right)\right)}{l_0^3}    
+\frac{\psi'\left(\frac{1-l_0}{2}\right)-\psi'\left(\frac{1+l_0}{2}\right)}
{4 l_0^2}\right]\right.\nonumber \\ &+&
\left.\delta_{l,l_0}\left[\frac{\frac{\pi^2 l}{2}+\psi(\frac{1}{2})
-\psi(\frac{1-2l}{2})-l\psi'(\frac{1-l}{2})}{4l^3}
\right]\right\}.\label{K-ll0res}
\end{eqnarray}
Here $\psi(z)$ is the digamma function, defined as
$\psi(z)=\Gamma'(z)/\Gamma(z)$, where $\Gamma(z)$ is the gamma
function.\cite{gr} For large arguments, $\psi(z)\sim \ln(z)$, which in
this context must be regarded as a constant, since it does not
contribute a power-law behaviour with non-zero exponent.  The
asymptotic behaviour of $K(l,l_0)$ is therefore to leading order given
by $K(l,l_0)\propto 1/l^2$.

\subsection{The $x-$ and $t$-dependence of the correlation functions}

The explicit calculation of $F$ is performed by bosonization of the chain
field operators,\cite{lutherpeschel,haldane} writing
\beq
\psi_{\alpha l s}(x,t)=
\frac{1}{\sqrt{2\pi\epsilon}}U^{\dagger}_{\alpha l s}
e^{-\alpha \varphi_{\alpha l s}(x,t)+i\alpha k_F x},
\eeq
where $\epsilon$ is a short-distance cutoff which should be sent to zero at
the end of all calculations. 
In principle, the phase field $\varphi$ contains both finite modes 
$(n_x\neq 0)$ and zero modes ($n_x=0$, corresponding to number operators), 
but the zero modes may be neglected in the thermodynamic limit.\cite{voit}
Furthermore, it is easily seen that the ladder operators $U$ do not give rise 
to any minus signs in the expectation values we consider, because there is no 
need to reorder the $U$'s before eliminating them using 
$U_{\alpha l s}U^{\dagger}_{\alpha l s}=U^{\dagger}_{\alpha l s}
U_{\alpha l s}=1$, and they can therefore be neglected.
The phase field is $\varphi_{\alpha l s}(x,t)=
\sum_{\nu}\varphi^{\nu}_{\alpha l s}(x,t)$, where we have defined 
\beq
\varphi^{\nu}_{\alpha l s}(x,t)=\frac{\kappa_{\nu s}}{2\sqrt{2}N}
\sum_{\vec{n}}\frac{e^{-\epsilon |k_x|/2}}{n_x}\nu_{\alpha}(\vec{n},t)
e^{-ik_x x-ip_y l},
\label{phase} 
\eeq
where $\kappa_{\rho s}=1$, $\kappa_{\sigma s}=s$ with $s=1$ $(-1)$ for
spin up (down). The time dependence is given by 
$\nu_{\alpha}(\vec{n},t)=e^{i(H_0+H_B)t}\nu_{\alpha}(\vec{n})
e^{-i(H_0+H_B)t}$. From $\nu^{\dagger}_{\alpha}(\vec{n})=\nu_{\alpha}
(-\vec{n})$, we obtain $\varphi^{\dagger}_{\alpha l s}(x,t)
=-\varphi_{\alpha l s}(x,t)$ . This gives 
\begin{eqnarray}
F_{\beta,s_1 s_2}(l_1-l_2,x,t) &=& (2\pi\epsilon)^{-2} \exp{[-ik_F x(1+\beta)]}
\prod_{\nu=\rho,\sigma}\exp{[E^{\nu}_{\beta,s_1s_2}(l_1-l_2,x,t)]}, 
\label{F}\\
\exp{[E^{\nu}_{\beta,s_1s_2}(l_1-l_2,x,t)]} &=&
\langle | e^{\varphi^{\nu}_{+l_1 s_1}(x,t)}e^{\beta\varphi^{\nu}
_{- l_2 s_2}(x,t)}
e^{-\beta\varphi^{\nu}_{- l_2 s_2}(0,0)}e^{-\varphi^{\nu}
_{+ l_1 s_1}(0,0)}|\rangle.
\end{eqnarray}
Next, we perform the canonical transformation with the operator (\ref{cantr}).
Its effect on the spin and charge operators is given by
\beq
e^{S_{\nu}}\nu_{\alpha}(\vec{n},t)e^{-S_{\nu}}=\sum_{\lambda=\pm 1}
h^{\alpha\lambda}_{\nu}(\vec{n})\nu_{\lambda}(\vec{n})
e^{i\lambda k_x v_{\nu}(\vec{n})t},
\eeq
where
\beq
h_{\nu}^{\alpha,\alpha}(\vec{n}) = \cosh\xi_{\nu}(\vec{n}), \;\;\;
h_{\nu}^{\alpha,-\alpha}(\vec{n}) = -\sinh\xi_{\nu}(\vec{n}). 
\eeq
The canonical transformation will mix the operators for the 
'+' and '-' faces in the phase fields:
\beq
e^{S_{\nu}}\varphi^{\nu}_{\alpha l s}(x,t)e^{-S_{\nu}}=
\sum_{\lambda=\pm 1}\tilde{\varphi}^{\nu_{\lambda}}_{\alpha l s}(x,t),
\eeq
where
\begin{eqnarray}
\tilde{\varphi}_{\alpha l s}^{\nu_{\lambda}}(x,t)=
\frac{\kappa_{\nu s}}{2\sqrt{2}N}\sum_{\vec{n}}
\frac{e^{-\epsilon |k_x|/2}}{n_x}h_{\nu}^{\alpha\lambda}(\vec{n})  
\nu_{\lambda}(\vec{n}) 
e^{i\lambda k_x v_{\nu}(\vec{n})t-ik_x x-ip_y l}.
\end{eqnarray}
This gives 
\begin{eqnarray}
\exp{[E^{\nu}_{\beta,s_1 s_2}(l_1-l_2,x,t)]}=\prod_{\lambda=\pm 1}  
\langle G|
e^{\tilde{\varphi}_{+ l_1 s_1}^{\nu_{\lambda}}(x,t)}
e^{\beta\tilde{\varphi}_{- l_2 s_2}^{\nu_{\lambda}}(x,t)}
e^{-\beta\tilde{\varphi}_{- l_2 s_2}^{\nu_{\lambda}}(0,0)}
e^{-\tilde{\varphi}_{+ l_1 s_1}^{\nu_{\lambda}}(0,0)}
|G\rangle,
\label{expE}
\end{eqnarray}
where $|G\rangle\equiv \exp{(\sum_{\nu}S_{\nu})}|\rangle$ is the ground state
of $H_D$.

The calculation of the expectation value is done by writing the
$\tilde{\varphi}_{\alpha l s}^{\nu_{\lambda}}$ fields as a sum 
of two contributions, one over positive $n_x$ and one over negative $n_x$. 
One of the contributions will contain creation operators, the 
other will contain annihilation operators. By using the relations
\begin{eqnarray}
e^{A}e^{B} &=& e^{A+B}e^{[A,B]/2},  \label{commplus}\\
e^{A}e^{B} &=& e^{B}e^{A}e^{[A,B]}, \label{commult}
\end{eqnarray}
we can move the exponentials containing creation operators to the left and
the exponentials containing annihilation operators to the right. In doing so,
we will generate a string of c-numbers due to the reordering of the 
exponentials which we can take outside the expectation value brackets. Since 
the annihilation operators acting to the right and the creation operators 
acting to the left annihilate the ground state, the expectation value of the 
operators which are left inside the expectation value brackets is just 1. 

First, the  phase fields are split into the two different contributions,
\beq
\tilde{\varphi}_{\alpha l s}^{\nu_{\lambda}}(x,t)=
\tilde{\varphi}_{\alpha l s}^{\nu_{\lambda},+}(x,t)+
\tilde{\varphi}_{\alpha l s}^{\nu_{\lambda},-}(x,t),
\eeq
where the superscripts $+$ and $-$ denote creation and annihilation, 
respectively, and
\begin{eqnarray}
\tilde{\varphi}_{\alpha l s}^{\nu_{\lambda},\pm}(x,t)=
\pm \frac{\lambda \kappa_{\nu s}}{2\sqrt{2}N}\sum_{\vec{n}}\,\!'
\frac{e^{-\epsilon k_x/2}}{n_x}
h_{\nu}^{\alpha\lambda}(\vec{n})\nu_{\lambda}(\pm \lambda\vec{n})  
e^{\pm[ik_x v_{\nu}(\vec{n})t-i\lambda k_x x-i\lambda p_y l]}, 
\end{eqnarray}
where we also used the evenness of $\xi_{\nu}(\vec{n})$, and where the prime
after the summation sign indicates that only positive $n_x$ should be summed
over. Then we use (\ref{commplus}) to write each exponential in (\ref{expE}) as
a product of exponentials of creation and annihilation parts of the phase
field. For this, and further calculations, we will need the commutator
\beq
[\tilde{\varphi}_{\alpha l s}^{\nu_{\lambda},+}(x,t),\tilde{\varphi}
_{\alpha' l' s'}^{\nu_{\lambda},-}(x',t')]=
\frac{\kappa_{\nu s}\kappa_{\nu s'}}{4N}\sum_{\vec{n}}\,\!'
\frac{e^{-\epsilon k_x}}{n_x} 
h_{\nu}^{\alpha\lambda}(\vec{n})h_{\nu}^{\alpha'\lambda}(\vec{n})
e^{ik_x v_{\nu}(\vec{n})(t-t')-i\lambda[k_x(x-x')+p_y(l-l')]}.
\label{commut}
\eeq
Using (\ref{commplus}) on all exponentials in (\ref{expE})
we generate the factor $\exp{[C^{\nu}]}$ where
\begin{eqnarray}
C^{\nu}
&=&
-\frac{1}{2N}\sum_{\vec{n}}\,\!'
\frac{e^{-\epsilon k_x}}{n_x}\cosh 2\xi_{\nu}(\vec{n}),
\end{eqnarray}
where we also used $\kappa_{\nu s}^2=1$. Using (\ref{commult}) to move
all annihilation operators to the right in (\ref{expE}), we generate the
factor $\exp{[D^{\nu\lambda}_{\beta,s_1 s_2}(l_1-l_2,x,t)]}$, where
\begin{eqnarray}
\lefteqn{
D^{\nu\lambda}_{\beta,s_1 s_2}(l_1-l_2,x,t)=
\frac{1}{4N}\sum_{\vec{n}}\,\!'\frac{e^{-\epsilon k_x}}{n_x}}\nonumber \\ & &
[\beta\kappa_{\nu s_1}\kappa_{\nu s_2} \sinh 2\xi_{\nu}(\vec{n})
(1-e^{-ik_x v_{\nu}(\vec{n})t+i\lambda k_x x})\cos p_y(l_1-l_2) 
+\cosh 2\xi_{\nu}(\vec{n}) e^{-ik_x v_{\nu}(\vec{n})t+i\lambda k_x x}].
\end{eqnarray}
The total exponent is 
\begin{eqnarray}
\lefteqn{E^{\nu}_{\beta,s_1 s_2}(l_1-l_2,x,t)
=C^{\nu}+\sum_{\lambda=\pm 1}D_{\beta,s_1 s_2}^{\nu\lambda}(l_1-l_2,x,t)
= -\frac{1}{\pi}\int_0^{\pi} dp_y\int_0^{\infty}dk_x} \nonumber \\ & &
[\cosh 2\xi_{\nu}(\vec{n})
-\beta \kappa_{\nu s_1}
\kappa_{\nu s_2}\sinh 2\xi_{\nu}(\vec{n})\cos p_y(l_1-l_2)]
\frac{e^{-\epsilon k_x}}{k_x}\left(1-e^{-ik_x v_{\nu}
(\vec{n})t}\cos k_x x\right).
\label{totexp}
\end{eqnarray}
A momentum transfer cutoff must be imposed on the interactions. For fixed
$n_y$, the treatment is as for the 1D case. In the equation above, we express
the integral as a sum of a non-interacting and an interacting part, by writing
$[\cdots]=1+([\cdots]-1)$. In both parts, we then multiply the integrand by
$[e^{-\Lambda_{\nu}(n_y)k_x}+(1-e^{-\Lambda_{\nu}(n_y)k_x})]$ to obtain two 
terms for each of the two parts. (For simplicity, we take the cutoff
function $\Lambda_{\nu}(n_y)$ to be the same for both parts; since the
non-interacting part is independent of $\beta$, $s_1 s_2$ and $l_1-l_2$, this
implies that the cutoff function be independent of these variables.)
In the first term, the main 
contributions will come from $k_x\ll 1/\Lambda_{\nu}(n_y)$, where we can 
replace $\xi_{\nu}(\vec{n})\to \xi_{\nu}(n_x=0,n_y)\equiv \xi_{\nu}(n_y)$ and
$v_{\nu}(\vec{n})\to v_{\nu}(n_x=0,n_y)\equiv v_{\nu}(n_y)$. In the second
term, which can be neglected for the interacting part, the main contributions
come from $k_x\gg 1/\Lambda_{\nu}(n_y)$, so we can replace 
$v_{\nu}(\vec{n})\to v_0$ there. This gives
\begin{eqnarray}
E^{\nu}_{\beta,s_1 s_2}(l_1-l_2,x,t) &=&   
-\frac{1}{2}\ln\left(\frac{(\epsilon+iv_0 t)^2+x^2}{\epsilon^2}\right)
-\frac{1}{2\pi}\int_0^{\pi} dp_y\, \left\{\ln\left(\frac
{(\Lambda_{\nu}(n_y)+iv_{\nu}(n_y)t)^2+x^2}{(\Lambda_{\nu}(n_y)+iv_0 t)^2+x^2}
\right)\right. \nonumber \\ &+&\left.
A^{\nu}_{\beta,s_1 s_2}(n_y,l_1-l_2)
\ln\left(\frac{(\Lambda_{\nu}(n_y)+iv_{\nu}(n_y)t)^2+x^2}{\Lambda_{\nu}^2(n_y)}
\right)\right\},
\end{eqnarray}
where we have defined
\beq
A^{\nu}_{\beta,s_1 s_2}(n_y,l_1-l_2)=-1+\cosh 2\xi_{\nu}(n_y)-\beta 
\kappa_{\nu s_1}\kappa_{\nu s_2}
\sinh 2\xi_{\nu}(n_y)\cos p_y (l_1-l_2).
\label{Afunc}
\eeq
Introducing $n_y$-averaged velocities $\bar{v}_{\nu}$ and cutoffs
$\bar{\Lambda}_{\nu}$, and adding and subtracting terms in the integrand, 
we may write
\begin{eqnarray}
\lefteqn{\prod_{\nu=\rho,\sigma}
\exp{[E^{\nu}_{\beta,s_1 s_2}(l_1-l_2,x,t)]}} \nonumber \\ &=&\frac{\epsilon^2}
{(\epsilon+iv_0 t)^2+x^2}\prod_{\nu=\rho,\sigma}
\left(\frac{(\bar{\Lambda}_{\nu}+iv_0 t)^2+x^2}
{(\bar{\Lambda}_{\nu}+i\bar{v}_{\nu} t)^2+x^2}\right)^{1/2}
\left(\frac{\bar{\Lambda}_{\nu}^2}{(\bar{\Lambda}_{\nu}+i\bar{v}
_{\nu} t)^2+x^2}\right)
^{d^{\nu}_{\beta,s_1 s_2}(l_1-l_2)}  \nonumber \\ & & 
\exp\left[-\frac{1}{2\pi}\int_0^{\pi} dp_y\,\left\{
\ln\left(\left[\frac{(\Lambda_{\nu}(n_y)+iv_{\nu}(n_y)t)^2+x^2}
{(\Lambda_{\nu}(n_y)+iv_0 t)^2+x^2}\right]\biggl/\left[
\frac{(\bar{\Lambda}_{\nu}+i\bar{v}_{\nu}t)^2+x^2}
{(\bar{\Lambda}_{\nu}+iv_0 t)^2+x^2}\right]\right) \right.\right. \nonumber 
\\ &+& \left.\left.
A^{\nu}_{\beta,s_1 s_2}(n_y,l_1-l_2)
\ln\left(\left[\frac{(\Lambda_{\nu}(n_y)+iv_{\nu}(n_y)t)^2+x^2}
{\Lambda_{\nu}^2(n_y)}\right]\biggl/\left[\frac{(\bar{\Lambda}_{\nu}+
i\bar{v}_{\nu}t)^2+
x^2}{\bar{\Lambda}_{\nu}^2}\right]\right)\right\}
\right],
\label{expEfin}
\end{eqnarray}
where we have defined the charge and spin exponents
\beq
d^{\nu}_{\beta,s_1 s_2}(l_1-l_2) = \frac{1}{2\pi}\int_0^{\pi} dp_y\,
A^{\nu}_{\beta,s_1 s_2}(n_y,l_1-l_2).
\label{dexp}
\eeq
Summing up the results so far, the correlation functions 
(\ref{rcdw})-(\ref{rts}) are given by 
Eqs. (\ref{RitoKF})-(\ref{K-ll0res}), (\ref{F}), and 
(\ref{Afunc})-(\ref{dexp}).

\subsection{Correlation function exponents}
\label{correxp}

From Eqs. (\ref{rsdwx}) and (\ref{rsdwz}) we see that the correlation
functions for the $x$- and $z$-components of the SDW operator differ
only in the sign of $s_1 s_2$. Assuming spin-rotation invariance of
the underlying model, these correlation functions must be
identical. From Eq. (\ref{Afunc}) it is seen that this is obtained if
$\xi_{\sigma}(n_y)=0$, which gives $A^{\sigma}_{\beta,s_1
s_2}(n_y,l_0)=d^{\sigma}_{\beta,s_1 s_2}(l_0)=0$. Consequently, the
correlation functions become independent of $s_1 s_2$, and the $s_1
s_2$ subscript may therefore be omitted from now on.  The
spin-rotation invariance also has the effect of making the SS and TS
correlation functions identical. As for the 1D
case,\cite{voit2,schulz2} a more sophisticated (renormalization-group)
treatment is needed to lift the degeneracy between CDW and SDW
fluctuations, and between SS and TS fluctuations.

For $t=0$ it is seen that the factor $\exp[\;\cdots\;]$ on the rhs in
(\ref{expEfin}) is unity for $\nu=\sigma$, and for $\nu=\rho$ its
leading behaviour for $x\to\infty$ is independent of $x$. Thus the
exponent of the leading $x$-dependence of (\ref{expEfin}) is
$-2[1+d^{\rho}_{\beta}(l_0)]$. For asymptotic values of $t$ the
analysis is not so straightforward, since both the square roots and
the factors $\exp[\;\cdots\;]$ will depend on $x$ and $t$ to leading
order.  However, as long as all velocities involved are non-zero
(which we assume to be the case), it seems reasonable to neglect this
additional dependence, so that the following leading-order asymptotic
approximation may be used:
\beq
F_{\beta}(l_0,x,t) \propto \frac{e^{-ik_F x(1+\beta)}}{(\epsilon+iv_0 t)^2+x^2}
\left(\frac{\bar{\Lambda}_{\rho}^2}{(\bar{\Lambda}_{\rho}+i\bar{v}_{\rho}t)^2
+x^2}\right)^{d^{\rho}_{\beta}(l_0)}.
\label{Fasympt}
\eeq

We see that $F_{\beta}(l_0,x,t)$ has a form that resembles that of the
correlation functions of the 1D Luttinger model.  However, the
exponents $d^{\rho}_{\beta}(l_0)$ depend on the chain difference
$l_0$, and in general have a rather different form than the Luttinger
model exponents. The exceptions are the exponents for the equal-chain
terms $l_0=0$. We have, in our notation,\cite{fermivel}
\begin{eqnarray}
\begin{array}{rlllll}
\mbox{CDW/SDW:} & d^{\rho}_{\beta=+1}=-\frac{1}{2}(1-e^{-2\xi_{\rho}}) 
& \mbox{(1D)}, 
& & d^{\rho}_{\beta=+1}(l_0=0)=-\frac{1}{2\pi}\int_0^{\pi}dp_y\, 
(1-e^{-2\xi_{\rho}(n_y)}) & \mbox{(2D)} 
\label{cdwsdw0exp}\\
\mbox{SS/TS:}   & d^{\rho}_{\beta=-1}=-\frac{1}{2}(1-e^{+2\xi_{\rho}}) 
& \mbox{(1D)}, 
& & d^{\rho}_{\beta=-1}(l_0=0)=-\frac{1}{2\pi}\int_0^{\pi}dp_y\, 
(1-e^{+2\xi_{\rho}(n_y)}) & \mbox{(2D)} 
\label{ssts0exp}\\
\end{array}
\end{eqnarray}
It is seen that the form of these exponents is very similar in the 1D
and 2D case, the only difference being the averaging over $p_y$ in the
2D expressions. If this had been the set of exponents which determined
the leading behaviour of our correlation functions, then the analogy
to the Luttinger model would have been very strong indeed. However,
things are not quite that simple. Each correlation function is given
as a sum over contributions from different $l_0$. Since for all $l_0$,
$K(l,l_0)\propto 1/l^2$ for asymptotically large $l$, the asymptotic
behaviour of the various contributions differ only in their
$x$-$t$-dependence. Thus, for a given correlation function (i.e. a
given value of $\beta$) the leading asymptotic behaviour comes from
the value of $l_0$ which gives the smallest $d^{\rho}_{\beta}(l_0)$,
i.e. we must minimize
\beq
-\frac{\beta}{2\pi}\int_0^{\pi}dp_y\;\sinh 2\xi_{\rho}(n_y)\cos p_y l_0
\label{l0depexp}
\eeq
with respect to the integer $l_0$.  Assuming repulsive effective
interactions (i.e. $\xi_{\rho}(n_y)>0$), it is indeed true that
$d^{\rho}_{1}(l_0)$ is minimal for the term with $l_0=0$, which
therefore gives the slowest decay in the CDW/SDW correlation
functions.  However, for $d^{\rho}_{-1}(l_0)$ the term $l_0=0$
actually gives the {\em largest} exponent; the minimum exponent for
the SS/TS correlation functions comes from some non-zero $l_0$, and is
therefore not of the form in (\ref{ssts0exp}).  Consequently, the
relation between the leading exponents for the CDW/SDW and SS/TS
correlation functions is not as for the 1D Luttinger model, where
these exponents are related through simple scaling relations. Since
$d^{\rho}_{1}(0) < d^{\rho}_{-1}(l_0)$, regardless of $l_0$, we
conclude that the CDW/SDW fluctuations are dominant. This is the same
conclusion as for the Luttinger model for repulsive interactions,
although the detailed nature of the leading exponents is different.

Throughout this paper we have assumed that we are in the region of
parameter space where the square shape of the Fermi surface is stable
with respect to doping away from half-filling. This requires, when
spin-rotation invariance is invoked, that\cite{luther94}
\beq
\frac{1}{\pi}\int_0^{\pi}\sinh^2 \xi_{\rho}(n_y)(1-\cos p_y l_0)>2\;\;\;\;
\mbox{for all }l_0\neq 0
\label{squarestable}
\eeq
(since $l_0=1$ then minimizes the expression on the lhs of
(\ref{squarestable}), an equivalent requirement is that this
expression be $>2$ for $l_0=1$). Let us define $U=
U^{(\rho)}_{++}(0)a/4\pi v_0$ and
$\gamma=U^{(\rho)}_{+-}(0)/U^{(\rho)}_{++}(0)$. Having repulsive
interactions requires $U\geq 0$, $\gamma \geq 0$. Furthermore,
Eqs. (\ref{xiformula}) and (\ref{vformula}) are well-defined and real
only when $\gamma < 1+1/U$. In order to have (\ref{squarestable})
satisfied, there is a lower bound on $\gamma$, which is easily found
by assuming $\gamma<1$, and noting that for fixed $\gamma$, $\sinh^2
\xi_{\rho}(n_y)$ is maximal when $U\to\infty$, giving
\beq
\sinh^2 \xi_{\rho}(n_y)\approx \frac{1}{2}\left[\frac{1}{\sqrt{1-\gamma^2}}-
1\right].
\eeq
The condition (\ref{squarestable}) then requires
$\gamma>\gamma_{\rm{min}}=2\sqrt{6}/5\approx 0.9798$. There is also a
$\gamma$-dependent lower bound on $U$. The lowest such bound is found
numerically to occur at the parameter space boundary, for $U\approx
376$, corresponding to $\gamma=1+1/U\approx 1.00266.$ As $\gamma$ is
decreased from this value towards $\gamma_{\rm{min}}$, the lower bound
on $U$ is found numerically to increase towards infinity. A lower
bound $U\approx 376$ corresponds to $U^{(\rho)}_{++}(0)a/\pi
v_0\approx 1500$, an extremely high value. We do not know how large
the bare couplings must be in order to renormalize to such high
effective values. It may be that inclusion of scattering between
orthogonal faces could reduce the effective values needed, but that is
not clear.  In the region of parameter space where
(\ref{squarestable}) is satisfied, we find that $l_0=2$ minimizes
$d^{\rho}_{-1}(l_0)$ if $\gamma<1$ (the expression (\ref{l0depexp}) is
then negative), and $l_0=\infty$ minimizes $d^{\rho}_{-1}(l_0)$ if
$\gamma>1$ ((\ref{l0depexp}) is then 0). We do not have any physical
explanation for why these particular values of $l_0$ minimize the
SS/TS exponent in each particular case.

\section{SUMMARY AND CONCLUDING DISCUSSION}
\label{disc}

We have evaluated various ground state correlation functions for a 2D
bosonic Hamiltonian with spin-charge separation, introduced 
in Ref. \onlinecite{luther94}. The model was arrived at by mapping the
problem of 2D electrons on a square lattice with nearest-neighbor
hopping and Hubbard-like interactions on a square Fermi surface onto
two orthogonal sets of 1D chains. In the presence of interactions, it
was shown in Ref. \onlinecite{luther94} that single-particle hopping
between parallel chains was irrelevant; furthermore, for sufficiently
strong interactions, the square shape of the Fermi surface was
preserved even away from half-filling. The bosonic Hamiltonian
resulted from neglecting interactions between orthogonal chains and
parallel-chain interactions of non-bosonic form.

In the mapping used, the physical 2D field operator is written as a
sum of field operators on the 1D chains. It is the 2D field operators
which enter in the definitions of the correlation functions. As a
consequence, the resulting correlation function is a sum of many
terms, each of which takes the form of a Luttinger-model correlation
function in the time direction and the spatial direction parallel to
the chains. The chain indices enter in the prefactors and in the
exponents of the Luttinger-model correlation functions. In order to
find the leading behaviour of the correlation function for large times
and distances, one must identify the term with the smallest decay
exponent by minimizing the exponent with respect to the chain
separation.  The leading exponents were found to differ in form from
the Luttinger model exponents due to the 2D nature of the
problem. Specifically, the simple scaling relations valid for the
Luttinger model exponents were not valid for our model.
  
The fluctuations we considered were $2k_F$ charge-density wave and
spin-density wave, $s$-wave singlet superconductivity, and triplet
superconductivity. The condition of spin-rotation invariance fixed the
spin part of the correlation functions, making only the charge part
non-trivial.  The CDW and SDW correlation functions turned out to be
degenerate, and so did the SS and TS correlation functions. As for the
Luttinger model, the CDW/SDW fluctuations were dominant for effective
repulsive interactions.

One may ask whether our conclusion concerning the absence of divergent
superconducting fluctuations for repulsive interactions holds for
pairing in {\it any} symmetry channel. In view of the Kohn-Luttinger
effect,\cite{kohnlutt} one would expect to see $T=0$ superconductivity
in some, albeit possibly very high, angular momentum channel. It may
be that possible superconductivity is either hiding in higher
spin-triplet channels, or in higher singlet-channels, not considered
in this paper. Since $O_{SS}(\vec{r})$ is a local object, it does not
contain a $d_{x^2-y^2}$-component, and consequently our results do not
rule out the possibility of a superconductive instability in this
channel.

\acknowledgements A.S. acknowledges support from the Norwegian
Research Council under Grants No. 110566/410 and
No. 110569/410. J.O.F. was supported by a University Fellowship at
NTNU. J.O.F. and A.S. thank NORDITA for its hospitality during several
visits, when part of this work was done.



\begin{thebibliography}{99}
\bibitem{luther94a} Note that, effectively, the operator dimensions for 
backscattering 
and Umklapp terms, including fermion operators that are non-local in chain 
indices, 
have already been obtained for the $N$-chain problem, see Section V of Ref. 
\onlinecite{luther94}.
\bibitem{fermivel} The quantity $\xi_{\rho}(0)$ appearing in the formula for
the 1D exponents should be evaluated from (\ref{xiformula}) with $v_0$ 
replaced by $v_F$.
\end{thebibliography}
\end{document}